\documentclass[twocolumn,aps,pr,superscriptaddress,preprintnumbers,nofootinbib,10pt]{revtex4-2}
\usepackage{amsmath,amssymb}
\usepackage[dvipdf,dvips]{graphicx}
\usepackage{color}
\usepackage{hyperref}
\usepackage{url}
\usepackage{slashed}
\usepackage{subfigure}
\usepackage[usenames,dvipsnames]{xcolor}
\usepackage{amsmath}
\usepackage{amsfonts}
\usepackage{float} 
\usepackage{amssymb}
\usepackage{epsfig}
\usepackage{graphics}
\usepackage{euscript}
\usepackage{slashed}
\usepackage{epstopdf}
\usepackage[utf8]{inputenc}
\allowdisplaybreaks
\usepackage{palatino}
\usepackage[normalem]{ulem}
\usepackage{pifont}
\usepackage{dsfont}
\usepackage{MnSymbol}
\usepackage{verbatim}
\usepackage{graphicx}
\usepackage{latexsym}
\usepackage{mathrsfs}
\usepackage{amsmath}
\usepackage{amssymb}
\usepackage{babel}

\hypersetup{
colorlinks=true,
citecolor=blue,
citebordercolor=red,
linktoc=all,
linkcolor=blue,
urlcolor=blue
}

\def \and{\textmd{and}}

\begin{document}

\title{Local gauge-invariant vector operators in the adjoint $SU(2)$ Higgs model}

\author{Giovani Peruzzo\,\href{https://orcid.org/0000-0002-4631-5257}{\usebox{\ORCIDicon}}}\email{gperuzzofisica@gmail.com}
\affiliation{Instituto de F\'isica, Universidade Federal Fluminense, Campus da Praia Vermelha, Av. Litor\^anea s/n, 24210-346, Niter\'oi, RJ, Brazil}

\graphicspath{{./figs/}}
\newbox{\ORCIDicon}
\sbox{\ORCIDicon}{\large
	\includegraphics[width=0.8em]{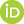}}

\begin{abstract}
In this work, we scrutinize local gauge-invariant vector operators of dimension four in the adjoint 
$SU(2)$ Higgs model, which are candidates for interpolating fields of the fundamental excitations of the model due to the so-called FMS mechanism. We use the equations of motion and the properties of the BRST operator to derive a Ward identity that allows us to determine whether a given operator can propagate. To corroborate this analysis, we explicitly compute the two-point function of the non-propagating operator at the one-loop level.

\end{abstract}

\maketitle

\section{Introduction}

The Higgs mechanism \cite{Higgs:1964pj,Englert:1964et,Guralnik:1964eu} plays a crucial role in the Standard Model of particle physics, being responsible for giving masses to the electroweak gauge bosons while preserving perturbative unitarity and renormalizability. This enormous phenomenological success has propelled the study of Higgs models beyond the perturbative level. In particular, significant results have emerged from the lattice formulation, which provides strong evidence, for instance, for the existence of confining phases, see the pioneering work of Fradkin and Shenker \cite{Fradkin:1978dv}. More recently, lattice studies have devoted special attention to the characterization of the physical spectrum using gauge-invariant operators, see Refs.~\cite{Maas:2016qpu,Torek:2017czn,Maas:2017xzh,Maas:2018ska,Torek:2018qet,Maas:2018xxu,Maas:2020kda,Maas:2022gdb,Maas:2017wzi}. Thanks to the so-called FMS (Fröhlich–Morchio–Strocchi) mechanism \cite{Frohlich:1980gj,Frohlich:1981yi}, it is possible to associate local gauge-invariant operators with the fundamental excitations of Higgs models, such as the Higgs boson and the gauge bosons. This special feature of Higgs models is particularly appealing in the lattice formulation. Due to Elitzur’s theorem~\cite{Elitzur:1975im}, expectation values of gauge-variant operators vanish in the absence of gauge fixing. On the lattice, where gauge fixing is not required, gauge-invariant operators are therefore the natural observables. In continuum space-time, on the other hand, gauge fixing is required; nevertheless, gauge-invariant operators remain of particular interest, since their Green’s functions are gauge independent and possess positive spectral representations, see Refs.~\cite{Kugo:1979gm,Dudal:2019aew,Dudal:2019pyg,Dudal:2020uwb}. \par  

In Ref.~\cite{Afferrante:2019vsr}, the authors computed, on the lattice, the two-point function of the lattice version of the gauge-invariant operator
\begin{equation}
	\frac{1}{\partial^2}\partial_\nu \left(\Phi^a F_{\nu\mu}^a\right) \label{eq:Maas_op}
\end{equation}
in the adjoint $SU(2)$ Higgs model, also known as a simplified version of the Georgi-Glashow model  \cite{Georgi:1972cj}. Corroborating the previous analysis of \cite{Lee:1985yi}, they obtained results indicating that this operator is associated with an Abelian massless vector mode. Within standard perturbation theory, this result can be understood by expanding $\Phi^a$ around one of the minima of the Higgs potential and noting that the leading contribution to this operator corresponds to the transverse component of the gauge field associated with the unbroken gauge-group generator. Since massless vector modes are identified with photons in grand unified theories, this provides a natural motivation to study such an operator, which is present in any adjoint Higgs model. \par

In this work, we analyze other gauge-invariant vector operators in the same adjoint $SU(2)$ Higgs model that share the same properties as the aforementioned operator, but formulated in continuum space-time within a gauge-fixed framework. We give particular attention to the local dimension-four operator
\begin{equation}
	O_\mu=\Phi^a D_{\nu}^{ab} F_{\nu\mu}^a, \label{eq:operator_vector}
\end{equation}
which has the same canonical dimension as the local version of Eq.~\eqref{eq:Maas_op}, namely 
$\partial_\nu \left(\Phi^a F_{\nu\mu}^a\right)$. From the renormalization point of view, we expect such an operator to appear in any case. However, although these two operators are equivalent at the linear level, we show that the second one does not propagate. We first demonstrate this result using the equations of motion and the properties of the BRST operator, and subsequently verify it by explicitly computing the connected two-point Green’s function $\langle O_{\mu}\left(x\right)O_{\nu}\left(y\right)$ up to one loop. These results establish the local version of Eq.~\eqref{eq:Maas_op} as the only local gauge-invariant vector operator of dimension four capable of describing the massless vector mode of the adjoint $SU(2)$ Higgs model. \par

The article is organized as follows. In Section~\ref{sec:Higgs action in the adjoint representation}, we briefly present the 
$SU(2)$ Higgs model in the adjoint representation. In Section~\ref{sec:Gauge fixing and BRST symmetry}, we introduce the gauge-fixing term and the BRST transformations. In Section~\ref{sec:Gauge-invariant vector operators}, we list the local gauge-invariant vector operators of dimension four that exist in the adjoint Higgs model. In Section~\ref{sec:Two-point correlation function}, we present the results for the two-point Green’s function of the operator~\eqref{eq:operator_vector}. In Section~\ref{sec:Other operators and Ward identities}, we extend the action to include gauge-invariant vector operators of dimension four and a non-linear BRST symmetry; as a result, we derive an enlarged set of Ward identities. Finally, in Section~\ref{sec:Conclusions}, we present our conclusions.

\section{Higgs action in the adjoint representation}\label{sec:Higgs action in the adjoint representation}

Let $\Phi^a\left(x\right)$ and $A_{\mu}^a\left(x\right)$, $a=1,2,3$, be a real scalar field and a gauge field, respectively. The Euclidean $SU\left(2\right)$ Higgs model in the adjoint representation
is defined by the following action:
\begin{widetext}
\begin{eqnarray}
	S_{\textrm{Higgs}} & = & \int d^{4}x\left[\frac{1}{4}F_{\mu\nu}^{a}F_{\mu\nu}^{a}+\frac{1}{2}\left(D_{\mu}\Phi\right)^{a}\left(D_{\mu}\Phi\right)^{a}+\frac{\lambda}{8}\left(\Phi^{a}\Phi^{a}-v^{2}\right)^{2}\right],
\end{eqnarray}	
\end{widetext}
where the field strength tensor and the covariant derivative are given
by
\begin{eqnarray}
	F_{\mu\nu}^{a} & = & \partial_{\mu}A_{\nu}^{a}-\partial_{\nu}A_{\mu}^{a}+g\varepsilon^{abc}A_{\mu}^{b}A_{\nu}^{c},\nonumber \\
	D_{\mu}^{ab} & = & \delta^{ab}\partial_{\mu}-g\varepsilon^{abc}A_{\mu}^{c},
\end{eqnarray}
respectively, where $\varepsilon^{abc}$ ($\varepsilon^{123}$) is
the Levi-Civita symbol. Notice that in this action there are three
parameter, namely, the gauge coupling ($g$), the quartic coupling
($\lambda$) and the massive parameter ($v$), which is related to
the vacuum expectation value (vev) of the scalar field. This action
is invariant under the following gauge transformations:
\begin{eqnarray}
	\Phi'{}^{a} & = & R^{ab}\Phi^{b},\nonumber \\
	A'_{\mu} & = & RA_{\mu}R^{-1}+\frac{1}{ig}\left(\partial_{\mu}R\right)R^{-1},
\end{eqnarray}
where $R=\exp\left(-ig\omega^{a}T^{a}\right)$ is the adjoint representation of an element of $SU\left(2\right)$ group. 

The classical potential
\begin{eqnarray}
	V\left(\Phi\right) & = & \frac{\lambda}{8}\left(\Phi^{a}\Phi^{a}-v^{2}\right)^{2}
\end{eqnarray}
has the typical Mexican hat shape, meaning that the vev of $\Phi^{a}$
does not vanish. Here we pick up the minima of $V\left(\Phi\right)$
in the third direction, \emph{i.e.},
\begin{eqnarray}
	\Phi_{o}^{a} & = & v\delta^{a3},
\end{eqnarray}
and expand $\Phi^{a}$ around it as follows:
\begin{eqnarray}
	\Phi^{a}\left(x\right) & = & v\delta^{a3}+H^{a}\left(x\right),
\end{eqnarray}
where $H^{a}$ is the truly propagating field. Rewriting the Higgs
action by using this parametrizations, we obtain
\begin{eqnarray}
	S_{\textrm{Higgs}} & = & \int d^{4}x\left[\frac{1}{4}F_{\mu\nu}^{a}F_{\mu\nu}^{a}+\frac{1}{2}\partial_{\mu}H^{\alpha}\partial_{\mu}H^{\alpha}+\frac{1}{2}\partial_{\mu}H^{3}\partial_{\mu}H^{3}\right.\nonumber \\
	&  & _{-}gv\varepsilon^{\alpha\beta}A_{\mu}^{\beta}\partial_{\mu}H_{\mu}^{\alpha}-gv\varepsilon^{abc}A_{\mu}^{c}H^{b}\partial_{\mu}H^{a}+\frac{g^{2}v^{2}}{2}A_{\mu}^{\alpha}A_{\mu}^{\alpha}\nonumber \\
	&  & -g^{2}v\varepsilon^{\alpha\beta}\varepsilon^{\alpha de}A_{\mu}^{\beta}A_{\mu}^{e}H^{d}+\frac{g^{2}}{2}\varepsilon^{abc}\varepsilon^{ade}A_{\mu}^{c}H^{b}A_{\mu}^{e}H^{d}\nonumber \\
	&  & \left.+\frac{\lambda v^{2}}{2}H^{3}H^{3}+\frac{\lambda v}{2}H^{3}H^{a}H^{a}+\frac{\lambda}{8}H^{a}H^{a}H^{b}H^{b}\right],
\end{eqnarray}
where we denote the components 1 and 2 of the gauge group by a Greek
letter, and $\varepsilon^{\alpha\beta}$ ($\varepsilon^{12}$) is
the two-dimensional Levi-Civita symbol. Notice that the components
$A_{\mu}^{1}$, $A_{\mu}^{2}$ and $H^{3}$ have acquired masses proportional
to the vev of $\Phi^{a}$, namely,
\begin{eqnarray}
	m & = & gv
\end{eqnarray}
and
\begin{eqnarray}
	m_{h} & = & \sqrt{\lambda}v,
\end{eqnarray}
respectively. Therefore, $H^{3}$ plays the role of the massive Higgs
field, whereas $H^{\alpha}$ is the would-be Goldstone field. This
picture is consistent with the fact that the generators $T^{1}$ and
$T^{2}$ ($\left(T^{c}\right)^{ab}=-i\varepsilon^{abc}$) are broken
by $\Phi_{o}$. The vacuum is only invariant under the Abelian subgroup
generated by $T^{3}$. 

\section{Gauge fixing and BRST symmetry}\label{sec:Gauge fixing and BRST symmetry}

To quantize the Higgs model, we must fix the gauge. We achieve this
by using the BRST quantization \cite{Becchi:1974md,Becchi:1974xu,Becchi:1975nq}. The BRST transformations of $A_{\mu}$
and $\Phi^{a}$ are infinitesimal gauge transformations in which $\omega^{a}\rightarrow c^{a}$,
where $c^{a}$ is the ghost field. Since $\delta A_{\mu}^{a}=-D_{\mu}\omega^{a}$
and $\delta\Phi^{a}=g\varepsilon^{abc}\omega^{b}\Phi^{c}$, thus
\begin{eqnarray}
	sA_{\mu}^{a} & = & -D_{\mu}c^{a}\nonumber \\
	sH^{3} & = & g\varepsilon^{\alpha\beta}c^{\alpha}H^{\beta},\nonumber \\
	sH^{\alpha} & = & gv\varepsilon^{\alpha\beta}c^{\beta}+g\varepsilon^{\alpha\beta}c^{\beta}H^{3}-g\varepsilon^{\alpha\beta}c^{3}H^{\beta}, \label{eq:brst_1}
\end{eqnarray}
where $s$ is the BRST operator. In addition to these fields, we have
the antighost and the Nakanishi-Lautrup fields- $\overline{c}^{a}$
and $b^{a}$, respectively- whose BRST transformations are
\begin{eqnarray}
	sc^{a} & = & \frac{g\varepsilon^{abc}}{2}c^{b}c^{c},\nonumber \\
	s\overline{c}^{a} & = & ib^{a},\nonumber \\
	sb^{a} & = & 0. \label{eq:brst_2}
\end{eqnarray}
These transformations imply that $s$ is nilpotent, \emph{i.e.},
\begin{eqnarray}
	s^{2} & = & 0.
\end{eqnarray}

The gauge fixing, $S_{gf}$, is a BRST-exact term, \emph{i.e.}, $S_{gf}=s\left(\dots\right)$, and the gauge condition we choose to impose over $A_{\mu}$ is given
by the equation of motion of the Nakanishi-Lautrup field. Usually,
we employ the renormalizable 't Hooft gauge \cite{tHooft:1971qjg}, also known as $R_{\xi}$
gauge, which is defined by
\begin{widetext}
\begin{eqnarray}
	S_{gf} & = & s\int d^{4}x\left[\overline{c}^{\alpha}\left(-\frac{\xi}{2}ib^{\alpha}+\partial_{\mu}A_{\mu}^{\alpha}-\xi gv\varepsilon^{\alpha\beta}H^{\alpha}\right)+\overline{c}^{3}\left(-\frac{\hat{\xi}}{2}ib^{3}+\partial_{\mu}A_{\mu}^{3}\right)\right],
\end{eqnarray}	
\end{widetext}
where $\xi$ and $\hat{\xi}$ are gauge parameters. Since we are interested
in Green's functions of gauge-invariant operators, which are gauge
independent, it suffices to consider the simplest case that corresponds
to the Landau gauge, where $\xi=\hat{\xi}=0$. Therefore, from now
on, we consider
\begin{eqnarray}
	S_{gf} & = & s\int d^{4}x\left[\overline{c}^{a}\partial_{\mu}A_{\mu}^{a}\right]\nonumber \\
	& = & \int d^{4}x\left[ib^{a}\partial_{\mu}A_{\mu}^{a}+\overline{c}^{a}\partial_{\mu}D_{\mu}^{ab}c^{b}\right].
\end{eqnarray}
Besides the absence of a gauge parameter, the Landau gauge also preserves
the global gauge symmetry of the theory.

From the gauge invariance of $S_{\textrm{Higgs}}$ and the nilpotency
property of $s$, we can easily demonstrate that the total action
$S_{\textrm{Higgs}}+S_{gf}$ is BRST-invariant. At the quantum level, this is the prominent symmetry, which allows us to define the physical subspace of states through its associated conserved charge.

\section{Gauge-invariant vector operators}\label{sec:Gauge-invariant vector operators}

As we previously mentioned, Afferrante et al (2019) \cite{Afferrante:2019vsr} analyzed the gauge-invariant
vector operator \eqref{eq:Maas_op}
on the lattice, which due to the Higgs mechanism has a linear term
proportional to $A_{\mu}^{3}$:
\begin{gather*}
	\frac{1}{\partial^{2}}\partial_{\nu}\left(\Phi^{a}F_{\nu\mu}^a\right)=v\left(\delta_{\mu\nu}-\frac{\partial_{\mu}\partial_{\nu}}{\partial^{2}}\right)A_{\mu}^{3}+\textrm{non-linear terms}.
\end{gather*}
Therefore, the two-point Green's function of $\frac{1}{\partial^{2}}\partial_{\nu}\left(\Phi^{a}F_{\nu\mu}\right)$
has a pole at $p^{2}=0$, which corresponds to the massless gauge
boson associated with $A_{\mu}^{3}$. The lattice simulations of Ref.
\cite{Afferrante:2019vsr} indicate that this result is also true at the non-perturbative
regime.

Here, we prefer to consider the local equivalent version of the aforementioned
vector operator, namely,
\begin{gather}
	V_\mu=\partial_{\nu}\left(\Phi^{a}F^a_{\nu\mu}\right),
\end{gather}
which has dimension four. Notice that, due to the antisymmetry of $F_{\mu\nu}$, this operator is transverse. In addition to this operator, there
are  other two non-equivalent local gauge-invariant vector operators
with the same dimension:
\begin{gather}
	\frac{v}{2}\partial_{\mu}\left(\Phi^{a}\Phi^{a}\right) \label{eq:vec_scalar}
\end{gather}
and the operator \eqref{eq:operator_vector}. The first operator \eqref{eq:vec_scalar} is longitudinal and equivalent to the scalar operator
\begin{equation}
	O = \frac{1}{2}\Phi^{a}\Phi^{a}-\frac{v^2}{2},
\end{equation}
which reads
\begin{eqnarray}
	O & = & vH^{3}+\frac{1}{2}H^{a}H^{a}.
\end{eqnarray}
The linear term in $H^{3}$ implies that the two-point function of
$O$ has a pole at $p^{2}=-m_{h}^{2}$ , indicating
that this operator is an interpolating field of the Higgs particle state. The
second operator looks promising in order to describe a massless vector
boson, as
\begin{eqnarray}
	O_{\mu} & = & v\left(\partial^{2}\delta_{\mu\nu}-\partial_{\mu}\partial_{\nu}\right)A_{\nu}^{3}+\textrm{non-linear terms}.
\end{eqnarray}
Hence, at the linear level, $O_{\mu}$ and $\partial_{\nu}\left(\Phi^{a}F_{\nu\mu}\right)$
cannot be distinguished. However, in the next section we present an
argument based on the equations of motion that $O_{\mu}$ does not propagate.
In other words, we show that the two-point Green's function $\left\langle O_{\mu}\left(p\right)O_{\nu}\left(-p\right)\right\rangle $
is an analytical function of the momentum $p$. Therefore, $\partial_{\nu}\left(\Phi^{a}F_{\nu\mu}\right)$
seems to be the only interesting vector operator with dimension four
that can be associated to the massless gauge boson. 

\section{Two-point Green's function of $O_{\mu}$}\label{sec:Two-point correlation function}

\subsection{Results from the equations of motion}

In this section, we will present a formal proof that, in fact, $O_{\mu}$
does not propagate. The proof starts from the equation of motion of
$A_{\mu}^{a}$:
\begin{equation}
	\frac{\delta S}{\delta A_{\mu}^{a}} = -D_{\nu}^{ab}F_{\nu\mu}^{b}-g\varepsilon^{bca}\Phi^{c}D_{\mu}^{bd}\Phi^{d}-i\partial_{\mu}b^{a}+g\varepsilon^{bca}\left(\partial_{\mu}\overline{c}^{b}\right)c^{c},\label{eq:A_equation_of_motion}
\end{equation}
where 
\begin{eqnarray}
	S & = & S_{\textrm{Higgs}}+S_{gf}\label{eq:S_total}
\end{eqnarray}
is the gauge fixed Higgs action in the Landau gauge. Multiplying (\ref{eq:A_equation_of_motion}) by $\Phi^{a}$,
we obtain
\begin{eqnarray*}
	\Phi^{a}\frac{\delta S}{\delta A_{\mu}^{a}} & = & -\Phi^{a}D_{\nu}^{ab}F_{\nu\mu}^{b}-i\Phi^{a}\partial_{\mu}b^{a}+g\varepsilon^{bca}\Phi^{a}\left(\partial_{\mu}\overline{c}^{b}\right)c^{c}\\
	& = & -O_{\mu}-s\left(\Phi^{a}\partial_{\mu}\overline{c}^{a}\right),
\end{eqnarray*}
which implies that
\begin{eqnarray}
	O_{\mu} & = & -\Phi^{a}\frac{\delta S}{\delta A_{\mu}^{a}}-s\left(\Phi^{a}\partial_{\mu}\overline{c}^{a}\right).\label{eq:O_mu_expression}
\end{eqnarray}
This last equation (\ref{eq:O_mu_expression}) reveals that $O_{\mu}$,
at the classical level, is essentially a BRST-exact term, thereby,
it cannot play any relevant physical role. Furthermore, notice that $\Phi^a\frac{\delta S}{\delta A^a_\mu}$ is also BRST-invariant.

To verify the consequences of (\ref{eq:O_mu_expression}) at the quantum
level, let us analyze the two-point Green's function of $O_{\mu}$,
which is defined by
\begin{eqnarray}
	\left\langle O_{\mu}\left(x\right)O_{\nu}\left(y\right)\right\rangle  & = & N\int D\varphi O_{\mu}\left(x\right)O_{\nu}\left(y\right)e^{-S},
\end{eqnarray}
where $D\varphi \equiv DA_{\mu}^{a}DH^{a}Dc^{a}D\overline{c}^{a}Db^{a}$ denotes the integration over all fields 
and $N^{-1}=\int D\varphi\, e^{-S}$
is the normalization constant. Using (\ref{eq:O_mu_expression}) to
rewrite $O_{\mu}\left(x\right)$, we obtain that
\begin{widetext}
\begin{gather}
	\left\langle O_{\mu}\left(x\right)O_{\nu}\left(y\right)\right\rangle  = \left\langle \left(\Phi^{a}\frac{\delta S}{\delta A_{\mu}^{a}}\right)\left(x\right)\left(\Phi^b\frac{\delta S}{\delta A_\nu^b}\right)\left(y\right)\right\rangle +\left\langle s\left(\Phi^{a}\partial_{\mu}\overline{c}^{a}\right)\left(x\right)\left(\Phi^b\frac{\delta S}{\delta A_\nu^b}\right)\left(y\right)\right\rangle+ \left\langle\left(\Phi^a\frac{\delta S}{\delta A_\mu^a}\right)\left(x\right) s\left(\Phi^{b}\partial_{\nu}\overline{c}^{b}\right)\left(y\right)\right\rangle  \nonumber  \\
	+ \left\langle\left(\Phi^{a}\partial_{\mu}\overline{c}^{a}\right)\left(x\right) s\left(\Phi^{b}\partial_{\nu}\overline{c}^{b}\right)\left(y\right)\right\rangle \label{eq:OO_ext}
\end{gather}	
\end{widetext}
The last three terms in the RHS of \eqref{eq:OO_ext} are zero due to the BRST symmetry of the
theory, since they can be written as BRST-exact variations, \emph{i.e.},
\begin{gather}
	\langle s\left(\dots\right)\rangle =0.
\end{gather}
The first term in the RHS of \eqref{eq:OO_ext} can be rewritten as
\begin{widetext}
\begin{eqnarray}
	\left\langle \Phi^{a}\left(x\right)\frac{\delta S}{\delta A_{\mu}^{a}\left(x\right)}\Phi^{b}\left(y\right)\frac{\delta S}{\delta A_{\nu}^{b}\left(y\right)}\right\rangle  & = & -N\int D\varphi\Phi^{a}\left(x\right)\frac{\delta S}{\delta A_{\mu}^{a}\left(x\right)}\Phi^{b}\left(y\right)\frac{\delta}{\delta A_{\nu}^{b}\left(y\right)}\left(e^{-S}\right)\nonumber \\
	& = & N\int D\varphi\frac{\delta}{\delta A_{\nu}^{b}\left(y\right)}\left[\Phi^{a}\left(x\right)\frac{\delta S}{\delta A_{\mu}^{a}\left(x\right)}\Phi^{b}\left(y\right)\right]e^{-S}\nonumber \\
	& = & N\int D\varphi\Phi^{a}\left(x\right)\frac{\delta^{2}S}{\delta A_{\nu}^{b}\left(y\right)\delta A_{\mu}^{a}\left(x\right)}\Phi^{b}\left(y\right)e^{-S},
\end{eqnarray}	
\end{widetext}
where we used the property that an integral of a total functional
derivative vanishes. Therefore,
we obtain
\begin{equation}
	\left\langle O_{\mu}\left(x\right)O_{\nu}\left(y\right)\right\rangle  = \left\langle \Phi^{a}\left(x\right)\frac{\delta^{2}S}{\delta A_{\nu}^{b}\left(y\right)\delta A_{\mu}^{a}\left(x\right)}\Phi^{b}\left(y\right)\right\rangle .\label{eq:OO_1-1}
\end{equation}
Since the action is local, we expect $\frac{\delta^{2}S}{\delta A_{\nu}^{b}\left(y\right)\delta A_{\mu}^{a}\left(x\right)}$
to contain a Dirac delta function $\delta^{4}\left(x-y\right)$ whose
effect is to impose that $y=x$ . Indeed, we have that
\begin{widetext}
\begin{eqnarray}
	\Phi^{a}\left(x\right)\frac{\delta^{2}S}{\delta A_{\nu}^{b}\left(y\right)\delta A_{\mu}^{a}\left(x\right)}\Phi^{b}\left(y\right) & = & \Phi^{a}\left(x\right)\Phi^{b}\left(y\right)\left(\delta_{\mu\nu}D_{\alpha}^{ac}D_{\alpha}^{cb}-D_{\nu}^{ac}D_{\mu}^{cb}\right)\delta^{4}\left(x-y\right),
\end{eqnarray}	
\end{widetext}
meaning that in the momentum space $\left\langle O_{\mu}\left(x\right)O_{\nu}\left(y\right)\right\rangle $
is a local function of the external momenta. \par

The equation \eqref{eq:O_mu_expression} also has implications to other Green's functions of $O_{\mu}$. In special, for the n-point function, we have that
\begin{widetext}
\begin{equation}
	\langle O_{\mu_1}(x_1)\dots O_{\mu_n} (x_n)\rangle=\left(-1\right)^n(n-1)!\,\langle \Phi^{a_1}(x_1)\dots \Phi^{a_n}(x_n) \frac{\delta^n S}{\delta A_{\mu_n}^{a_n}(x_n)\dots A_{\mu_1}^{a_1}(x_1)} \rangle, \label{eq:n_point_function}
\end{equation} 	
\end{widetext}
which is also a non-propagating object due to the product of delta functions $\delta^4(x_1-x_2)\dots \delta^4(x_1- x_n)$ coming from the n\textsuperscript{th}-derivative of $S$. In particular, this function vanishes for $n>2$, see Section \ref{sec:Other operators and Ward identities}. In a similar way, more results can be derived for other Green's functions of $O_\mu$.

\subsection{Explicit one-loop results}

In Appendix \ref{One-loop corrections for}, we explicitly show all the one-loop contributions to
$\left\langle O_{\mu}\left(x\right)O_{\nu}\left(y\right)\right\rangle $
in the momentum space, which we define as 
\begin{equation}
	\left\langle O_{\mu}\left(x\right)O_{\nu}\left(y\right)\right\rangle  = \int\frac{d^{d}p}{\left(2\pi\right)^{d}}e^{-ip\cdot\left(x-y\right)}\left\langle O_{\mu}\left(p\right)O_{\nu}\left(-p\right)\right\rangle .
\end{equation}
To calculate these perturbative corrections, we employ Dimensional Regularization \cite{tHooft:1972tcz,Bollini:1972ui}, which preserves all symmetries of the model.
The result we obtain for the regularized function in $d$ dimensions
is 
\begin{widetext}
\begin{eqnarray}
	\left\langle O_{\mu}\left(p\right)O_{\nu}\left(-p\right)\right\rangle  & = &v^2(p^2\delta_{\mu\nu}-p_{\mu}p_{\nu})+ \delta_{\mu\nu}\left(-m_{h}^{2}+p^{2}\right)\chi\left(m_{h}^{2}\right)\nonumber \\
	& & -\left(-\frac{m_{h}^{2}}{d}\delta_{\mu\nu}+p_{\mu}p_{\nu}\right)\chi\left(m_{h}^{2}\right)+2m^{2}\delta_{\mu\nu}\frac{\left(d-1\right)^{2}}{d}\chi\left(m^{2}\right) \nonumber \\
	&  & +2\left(p^{2}\delta_{\mu\nu}-p_{\mu}p_{\nu}\right)\left[-\frac{3}{2}\chi\left(m_{h}^{2}\right)+\frac{m^{2}}{m_{h}^{2}}\left(d-1\right)\chi\left(m^{2}\right)\right], 
\end{eqnarray}	
\end{widetext}
which is indeed  a local function of the external momenta, since
\begin{eqnarray}
	\chi\left(M^{2}\right) & = & \int\frac{d^{d}k}{\left(2\pi\right)^{d}}\frac{1}{k^{2}+M^{2}}\nonumber \\
	& = & \frac{1}{\left(4\pi\right)^{\frac{d}{2}}}\Gamma\left(1-\frac{d}{2}\right)\left(M^{2}\right)^{\frac{d}{2}-1}\label{eq:xi_master}
\end{eqnarray}
is a momentum independent quantity. \par 
Notice that if $v=0$, then $m=m_h=0$, and consequently $\chi \left(0\right)=0$. Therefore, the one-loop corrections of $\langle O_\mu \left(x\right) O_\nu \left(y\right) \rangle$ vanish. Indeed, in this case $\Phi^a\left(x\right)=H^a\left(x\right)$, all propagators are massless, and there is only the seventh diagram of Figure 2 contributing to this function, which turns out to be zero.  

\section{Other operators and the Ward identities}\label{sec:Other operators and Ward identities}
So far, we have used the Eq. \eqref{eq:O_mu_expression} to derive the results  for the Green's functions of $O_\mu$, such as the Eq. \eqref{eq:n_point_function}. Therefore, one can argue that these results are a bit formal and might not hold after we consider the renormalization. To avoid this criticism, we show that is possible to rigorously derive \eqref{eq:n_point_function} from a Ward identity. \par

Instead of action $S$, given by Eq. \eqref{eq:S_total}, let us consider the extended action 
\begin{eqnarray}
	\Sigma & = & S+\int d^{4}x\left(\Omega_{\mu}O_{\mu}+\Xi_{\mu\nu}Q_{\mu\nu}+\rho_\mu V_\mu+JO\right) \nonumber \\
	& &+\int d^4 x \left( Z_\mu \Phi^a\partial_\mu \bar{c}^a+Y_\mu s(\Phi^a\partial_\mu\bar{c}^a)\right) \nonumber \\
	& & +\int d^4 x \left(K^a_\mu sA_\mu^a+L^a sc^a+R^a sH^a \right),
\end{eqnarray}
which includes the non-linear BRST transformations, see Eqs. \eqref{eq:brst_1} and \eqref{eq:brst_2}, as well as the operators $O_\mu$, $V_\mu$, $O$ and 
\begin{eqnarray}
	Q_{\mu\nu} & = & \Phi^{a}\left(\delta_{\mu\nu}D_{\lambda}^{ab}D_{\lambda}^{bc}-D_{\nu}^{ab}D_{\mu}^{bc}\right)\Phi^{c}.
\end{eqnarray}
These operators are coupled with the external sources $K_\mu^a$, $L^a$, $R^a$, $\Omega_\mu$, $\Xi_{\mu\nu}$, $\rho_\mu$ and $J$, which are all BRST singlets, \emph{i.e.},
\begin{gather}
	sK^a_\mu=sL^a=sR^a=s\Omega_\mu=s\Xi_{\mu\nu}=s\rho_\mu=sJ=0.
\end{gather}
The operator $s\left(\Phi^a\partial_\mu \bar{c}^a\right)$ is BRST-exact, thus, we introduce it and $\Phi^a\partial_\mu \bar{c}^a$ \emph{via} a  doublet of sources, namely, $Y_\mu$ and $Z_\mu$, such that,
\begin{eqnarray}
	sY_\mu &=& Z_\mu, \nonumber \\
	sZ_\mu&=&0.
\end{eqnarray}
Therefore, we still preserve the BRST-symmetry of the theory, \emph{i.e.},
\begin{gather}
	s\Sigma =0.
\end{gather}
The BRST invariance of $\Sigma$ can be expressed through the Slavnov-Taylor identity
\begin{gather}
\int d^4x \left(\frac{\delta \Sigma}{\delta K_\mu^a}\frac{\delta \Sigma}{\delta A_\mu^a}+\frac{\delta \Sigma}{\delta L^a}\frac{\delta \Sigma}{\delta c^a}+\frac{\delta \Sigma}{\delta R^a}\frac{\delta \Sigma}{\delta H^a} \right. \nonumber \\
\left.+ib^a\frac{\delta \Sigma}{\delta \bar{c}^a}+Z_\mu\frac{\delta \Sigma}{\delta Y_\mu}\right)=0. \label{eq:slavnov_taylor}
\end{gather}

Since we are working in the Landau gauge, then we have the typical Ward identities associated with the equations of motion of the Nakanishi-Lautrup field, the ghost field and the antighost field- see Ref. \cite{Piguet:1995er}- namely,
\begin{equation}
	\frac{\delta \Sigma}{\delta b^a}=i\partial_\mu A^a_\mu-i\partial_{\mu}\left(Y_\mu \Phi^a\right), \label{eq:b_equation}
\end{equation}
\begin{gather}
\int d^4 x \left( \frac{\delta \Sigma}{\delta c^a} -ig\varepsilon^{abc}\bar{c}^b\frac{\delta \Sigma}{\delta b^c}	\right) \nonumber \\
= g\varepsilon^{abc}\int d^4 x \left(K_\mu^b A_\mu^c+c^bL^c+R^b\Phi^c \right),
\end{gather}
\begin{equation}
\frac{\delta \Sigma}{\delta \bar{c}^a}+\partial_\mu \frac{\delta \Sigma}{\delta K_\mu^a}-\partial_\mu \left(Y_\mu \frac{\delta \Sigma}{\delta R^a}\right)=0,
\end{equation}
respectively. With the inclusion of these new operators, the theory also has the following Ward identities:
\begin{gather}
\Phi^{a}\frac{\delta\Sigma}{\delta A_{\mu}^{a}}+\frac{\delta\Sigma}{\delta\Omega_{\mu}}-\Omega_{\nu}\frac{\delta\Sigma}{\delta\Xi_{\nu\mu}}+\frac{\delta\Sigma}{\delta Y_{\mu}}-K_{\mu}^{a}\frac{\delta\Sigma}{\delta R^{a}}\nonumber \\
=-\left(\partial_{\nu}\rho_{\mu}-\partial_{\mu}\rho_{\nu}\right)\partial_{\nu}\frac{\delta\Sigma}{\delta J}-2\left(\partial^{2}\rho_{\mu}-\partial_{\mu}\partial_{\nu}\rho_{\nu}\right)\frac{\delta\Sigma}{\delta J}\nonumber \\
+\partial_{\mu}\left(\Omega_{\nu}\partial_{\nu}\frac{\delta\Sigma}{\delta J}\right)-\partial_{\nu}\left(\Omega_{\nu}\partial_{\mu}\frac{\delta\Sigma}{\delta J}\right)\nonumber \\
+2\partial_{\nu}\left[\left(\partial_{\nu}\Omega_{\mu}-\partial_{\mu}\Omega_{\nu}\right)\frac{\delta\Sigma}{\delta J}\right]\nonumber \\
-v^{2}\left(\partial^{2}\rho_{\mu}-\partial_{\mu}\partial_{\nu}\rho_{\nu}\right)+v^{2}\left(\partial^{2}\Omega_{\mu}-\partial_{\mu}\partial_{\nu}\Omega_{\nu}\right), \label{eq:A_ward_identity}
\end{gather}

  \begin{equation}
 	\int d^4 x \left(\frac{\delta \Sigma}{\delta H^3}-\lambda v \frac{\delta \Sigma}{\delta J}\right)-\frac{\partial \Sigma}{\partial v}=\int d^4 x vJ, \label{eq:H_identity}
 \end{equation}
and 
 \begin{equation}
 	\partial_\mu \frac{\delta \Sigma}{\delta \rho_\mu}=0. \label{eq:rho_identity}
 \end{equation}
 
All the Ward identities listed above are non-anomalous, see \cite{Piguet:1995er}. Therefore, at the quantum level, one can construct an effective action $\Gamma$ that satisfies the same functional equations. In particular, the quantum version of Eq.~\eqref{eq:A_ward_identity} allows us to derive the results \eqref{eq:OO_1-1} and \eqref{eq:n_point_function} in a rigorous and transparent way to all orders, since
\begin{equation}
	\langle O_{\mu_{1}}\left(x_1\right)\dots O_{\mu_{n}}\left(x_n\right)\rangle_{1\text{PI}}= \left.\frac{\delta^n \Gamma}{\delta \Omega_{\mu_{1}}\left(x_1\right)\dots \Omega_{\mu_{n}}\left(x_n\right)}\right|_{j=0},
\end{equation}
where $j=0$ indicates that all external sources are set to zero. Equation~\eqref{eq:rho_identity}, in turn, guarantees that the Green’s functions of $V_\mu$ are transverse to all orders.

\section{Conclusions}\label{sec:Conclusions}
The possibility of constructing gauge-invariant interpolating operators for the fundamental excitations of Higgs models has renewed interest in these models within lattice gauge theory. However, the case of the operator $O_\mu$ shows that care must be taken when selecting which gauge-invariant operators to consider, in particular operators of high dimensionality. Although, at the linear level, this operator appears to be equivalent to the transverse component of $A_\mu^3$, in fact, it does not propagate. We therefore conclude that the FMS mechanism alone is not sufficient to guarantee that a given gauge-invariant operator serves as an interpolating field for a particular fundamental excitation. Such an analysis must be supplemented by a concrete non-perturbative investigation, which can be carried out using the equations of motion and Ward identities, as presented here and in Ref.~\cite{Capri:2020ppe,Dudal:2021dec,Peruzzo:2024heb}, or by means of lattice simulations. \par

The major consequence of our results for $O_\mu$ is that $V_\mu$ is the only remaining candidate for a local gauge-invariant vector operator with this dimensionality that can serve as an interpolating field. Looking ahead to the rigorous renormalization of its Green’s functions, we note that 
$\partial_\mu O $ is ruled out by the transversality of $V_\mu$, see \eqref{eq:rho_identity}. Therefore, it suffices to consider the operator $O_\mu$ in an algebraic renormalization analysis \cite{Piguet:1995er} of the operator $V_\mu$. Moreover, since the Green’s functions of $O_\mu$ are local, we expect that their loop corrections can be removed by an appropriate choice of local counterterms.

\section*{Acknowledgments}
The author would like to thank the Brazilian agency FAPERJ for financial support. G. Peruzzo is a FAPERJ postdoctoral fellow in the PÓS DOUTORADO NOTA 10 program under the contracts E-26/205.924/2022 and E-26/205.925/2022.


\appendix

\onecolumngrid

\section{Feynman Rules}

\subsection{Propagators}

The propagators of the theory are determined by the quadratic part
of $S$- see Eq. (\ref{eq:S_total})- namely,
\begin{eqnarray}
	S_{quad} & = & \int d^{d}x\left[\frac{1}{4}\left(\partial_{\mu}A_{\nu}^{\alpha}-\partial_{\nu}A_{\mu}^{\alpha}\right)^{2}+\frac{m^{2}}{2}A_{\mu}^{\alpha}A_{\mu}^{\alpha}+\frac{1}{4}\left(\partial_{\mu}A_{\nu}^{3}-\partial_{\nu}A_{\mu}^{3}\right)^{2}\right.\nonumber \\
	&  & \left.+\frac{1}{2}\left(\partial_{\mu}H^{3}\right)^{2}+\frac{m_{h}^{2}}{2}H^{2}+\frac{1}{2}\left(\partial_{\mu}H^{\alpha}\right)^{2}{}_{-}gv\varepsilon^{\alpha\beta}A_{\mu}^{\beta}\partial_{\mu}H_{\mu}^{\alpha}+ib^{a}\partial_{\mu}A_{\mu}^{a}+\overline{c}^{a}\partial^{2}c^{a}\right].\label{eq:S_quad}
\end{eqnarray}
In terms of the generating functional, they are defined as
\begin{eqnarray}
	\left(\Delta_{\varphi\varphi}\right)\left(x-y\right) & = & \left.\frac{\delta^{2}Z_{0}\left[J_{\varphi}\right]}{\delta J_{\varphi}\left(y\right)\delta J_{\varphi}\left(x\right)}\right|_{J_{\varphi}=0}
\end{eqnarray}
where
\begin{eqnarray}
	Z_{0}\left[J_{\varphi}\right] & = & \frac{\int D\varphi\,e^{-\left(S_{quad}+\int d^{4}xJ_{\varphi}\varphi\right)}}{\int D\varphi\,e^{-S_{quad}}}.
\end{eqnarray}
Here, we adopt the following convention for propagators in momentum
space:
\begin{eqnarray}
	\left(\Delta_{\varphi\varphi}\right)\left(x-y\right) & = & \int\frac{d^{4}p}{\left(2\pi\right)^{4}}e^{-ip\cdot\left(x-y\right)}\left(\Delta_{\varphi\varphi}\right)\left(p\right).
\end{eqnarray}
Therefore, we obtain the following non-vanishing propagators:
\begin{gather}
	\left(\Delta_{AA}\right)_{\mu\nu}^{\alpha\beta}\left(p\right) = \frac{\delta^{\alpha\beta}T_{\mu\nu}\left(p\right)}{p^{2}+m^{2}},\qquad
	\left(\Delta_{AA}\right)_{\mu\nu}^{33}\left(p\right) = \frac{T_{\mu\nu}\left(p\right)}{p^{2}},\nonumber \\
	\left(\Delta_{HH}\right)^{\alpha\beta}\left(p\right) = \frac{\delta^{\alpha\beta}}{p^{2}},\qquad
	\left(\Delta_{HH}\right)^{33}\left(p\right)= \frac{1}{p^{2}+m_{h}^{2}},\nonumber \\
	\left(\Delta_{c\overline{c}}\right)^{ab}\left(p\right) = \frac{\delta^{ab}}{p^{2}},\qquad
	\left(\Delta_{Ab}\right)_{\mu}^{ab}\left(p\right) =\frac{\delta^{ab}p_{\mu}}{p^{2}},
\end{gather}
where 
\begin{equation}
	T_{\mu\nu}\left(p\right) = \delta_{\mu\nu}-\frac{p_{\mu}p_{\nu}}{p^{2}}, \qquad
	L_{\mu\nu}\left(p\right) = \frac{p_{\mu}p_{\nu}}{p^{2}},
\end{equation}
are the transverse and the longitudinal projectors, respectively. We represent
the gauge field propagator by a wavy line, the scalar propagator by
a straight line, and the ghost propagator by a dashed line, as indicated in Figure 1.

\begin{figure}
	\begin{centering}
		\includegraphics[scale=0.35]{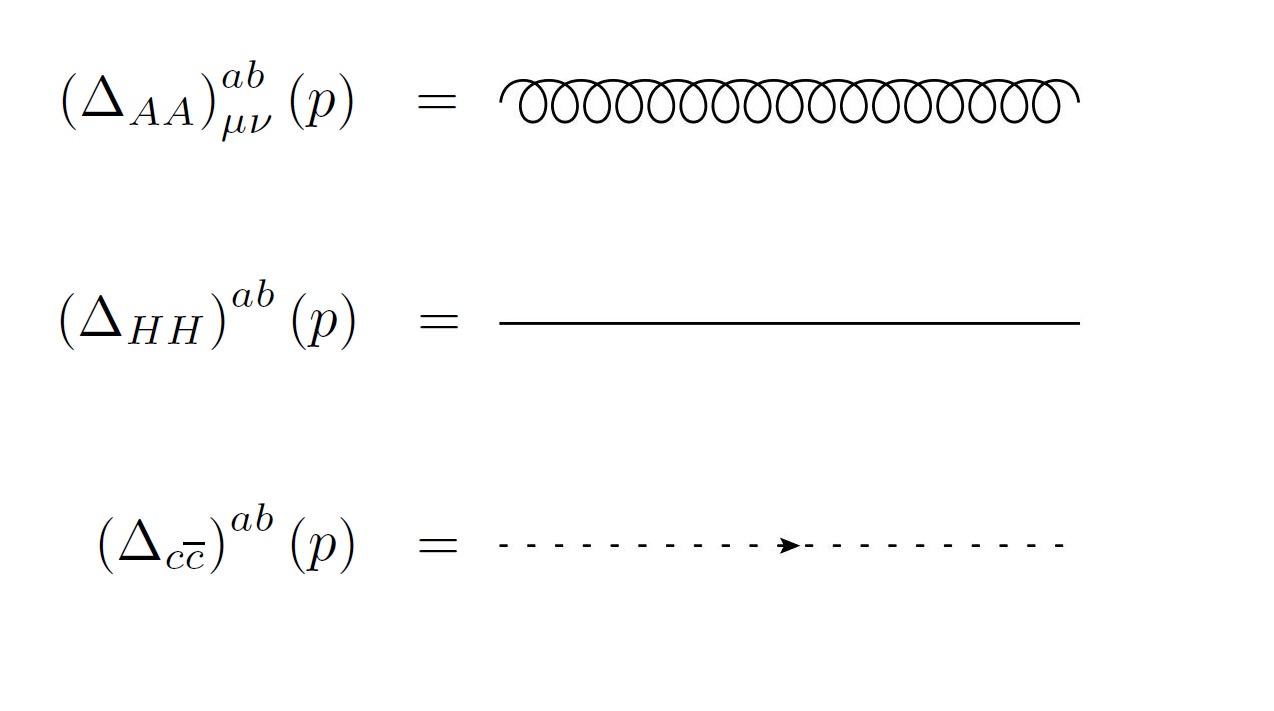}
		\par\end{centering}
	\caption{Diagrammatic representation of the relevant propagators. }\label{fig:propagators}
\end{figure}

\subsection{Vertices}

The interacting vertices are given by
\begin{eqnarray*}
	S_{I} & \equiv & S+\int d^{4}x\Omega_{\mu}\left(x\right)O_{\mu}\left(x\right)-S_{quad}\\
	& = & \int d^{4}x\left[g\varepsilon^{abc}\left(\partial_{\mu}A_{\nu}^{a}\right)A_{\mu}^{b}A_{\nu}^{c}+\frac{g^{2}}{4}\varepsilon^{abc}\varepsilon^{ade}A_{\mu}^{b}A_{\nu}^{c}A_{\mu}^{d}A_{\nu}^{e}\right.\\
	&  & -g\varepsilon^{acd}H^{c}A_{\mu}^{d}\partial_{\mu}H^{a}+vg^{2}\varepsilon^{a3d}\varepsilon^{aef}A_{\mu}^{d}H^{e}A_{\mu}^{f}+\frac{1}{2}g^{2}\varepsilon^{acd}\varepsilon^{aef}H^{c}A_{\mu}^{d}H^{e}A_{\mu}^{f}\\
	&  & \left.+\frac{\lambda v}{2}H^{3}H^{a}H^{a}+\frac{\lambda}{8}H^{a}H^{a}H^{b}H^{b}+g\varepsilon^{abc}A_{\mu}^{c}\partial_{\mu}\overline{c}^{a}c^{b}+\Omega_{\mu}\Phi^{a}D_{\nu}^{ab}F_{\nu\mu}^{b}\right]
\end{eqnarray*}
where $\Omega_{\mu}\left(x\right)$ is the external source that we
introduced to compute Green's functions of $O_{\mu}=\left(x\right)\Phi^{a}D_{\nu}^{ab}F_{\nu\mu}^{b}$.
In the momentum space, the vertices are defined by
\begin{eqnarray}
	\left(\Gamma_{\varphi_{i_{1}}\dots\varphi_{i_{n}}}\right)\left(p_{1},\dots,p_{n}\right) & = & -\int\prod_{k=1}^{n}d^{4}x_{k}e^{ip_{k}\cdot x_{k}}\left.\frac{\delta^{n}S_{I}}{\delta\varphi_{i_{1}}\left(x_{n}\right)\dots\delta\varphi_{i_{n}}\left(x_{1}\right)}\right|_{\varphi=0}. \label{eq:vertex_definition}
\end{eqnarray}
With the convention adopted for the propagators, this implies that
all momenta are flowing toward the vertices. 

The definition \eqref{eq:vertex_definition} leads to the following vertices of the Higgs model:
\begin{itemize}
	\item Three vectors vertex
\end{itemize}
\begin{eqnarray}
	\left(\Gamma_{AAA}\right)_{\alpha\beta\gamma}^{mnp}\left(p,q,r\right) & = & ig\varepsilon^{mnp}\left[\delta_{\alpha\gamma}\left(r_{\beta}-p_{\beta}\right)+\delta_{\alpha\beta}\left(p_{\gamma}-q_{\gamma}\right)+\delta_{\beta\gamma}\left(q_{\alpha}-r_{\alpha}\right)\right]\left(2\pi\right)^{4}\delta\left(p+q+r\right)
\end{eqnarray}

\begin{itemize}
	\item Four vectors vertex
\end{itemize}
\begin{eqnarray}
	\left(\Gamma_{AAAA}\right)_{\alpha\beta\gamma\delta}^{mnpq}\left(p,q,r,s\right) & = & -\left(2\pi\right)^{4}\delta^{4}\left(p+q+r+s\right)g^{2}\left[\left(\varepsilon^{amp}\varepsilon^{anq}+\varepsilon^{amq}\varepsilon^{anp}\right)\delta_{\alpha\beta}\delta_{\gamma\delta}\right.\nonumber \\
	&  & +\left(\varepsilon^{amn}\varepsilon^{apq}+\varepsilon^{amq}\varepsilon^{apn}\right)\delta_{\alpha\gamma}\delta_{\beta\delta}\nonumber \\
	&  & \left.+\left(\varepsilon^{amn}\varepsilon^{aqp}+\varepsilon^{amp}\varepsilon^{aqn}\right)\delta_{\alpha\delta}\delta_{\beta\gamma}\right]
\end{eqnarray}

\begin{itemize}
	\item Vector-anti-ghost-ghost vertex
\end{itemize}
\begin{eqnarray}
	\left(\Gamma_{A\overline{c}c}\right)_{\alpha}^{mnp}\left(p,q,r\right) & = & -ig\varepsilon^{npm}q_{\alpha}\left(2\pi\right)^{4}\delta^{4}\left(p+q+r\right)
\end{eqnarray}

\begin{itemize}
	\item Vector-vector-scalar vertex
\end{itemize}
\begin{eqnarray}
	\left(\Gamma_{AAH}\right)_{\alpha\beta}^{mnp}\left(p,q,r\right) & = & -vg^{2}\left(\varepsilon^{a3m}\varepsilon^{apn}+\varepsilon^{a3n}\varepsilon^{apm}\right)\delta_{\alpha\beta}\left(2\pi\right)^{4}\delta^{4}\left(p+q+r\right)
\end{eqnarray}

\begin{itemize}
	\item Vector-scalar-scalar vertex
\end{itemize}
\begin{eqnarray}
	\left(\Gamma_{AHH}\right)_{\alpha}^{mnp}\left(p,q,r\right) & = & g\varepsilon^{mnp}i\left(q_{\alpha}-r_{\alpha}\right)\left(2\pi\right)^{4}\delta^{4}\left(p+q+r\right)
\end{eqnarray}

\begin{itemize}
	\item Vector-vector-scalar-scalar vertex
\end{itemize}
\begin{eqnarray}
	\left(\Gamma_{AAHH}\right)_{\alpha\beta}^{mnpq} & = & -g^{2}\left(\varepsilon^{apm}\varepsilon^{aqn}\delta_{\alpha\beta}+\varepsilon^{aqm}\varepsilon^{apn}\right)\delta_{\alpha\beta}\left(2\pi\right)^{4}\delta^{4}\left(p+q+r+s\right)
\end{eqnarray}

\begin{itemize}
	\item Three scalars vertex
\end{itemize}
\begin{eqnarray}
	\left(\Gamma_{HHH}\right)^{mnp}\left(p,q,r\right) & = & -\lambda v\left(\delta^{m3}\delta^{np}+\delta^{mp}\delta^{n3}+\delta^{mn}\delta^{p3}\right)\left(2\pi\right)^{4}\delta^{4}\left(p+q+r\right)
\end{eqnarray}

\begin{itemize}
	\item Four scalars vertex
\end{itemize}
\begin{eqnarray}
	\left(\Gamma_{HHHH}\right)^{mnpq} & = & -\lambda\left(\delta^{mn}\delta^{pq}+\delta^{mp}\delta^{nq}+\delta^{mq}\delta^{np}\right)\left(2\pi\right)^{4}\delta^{4}\left(p+q+r+q\right)
\end{eqnarray}

In addition to these vertices, we also have those ones with the external
source $\Omega_{\mu}$:
\begin{itemize}
	\item Vector vertex
\end{itemize}
\begin{eqnarray}
	\left(\Gamma_{A}^{\Omega}\right)_{\alpha}^{m}\left(p\right) & = & v\delta^{m3}\left(\delta_{\alpha\beta}p^{2}-p_{\alpha}p_{\beta}\right)\tilde{\Omega}_{\beta}\left(p\right)
\end{eqnarray}

\begin{itemize}
	\item Two vectors vertex
\end{itemize}
\begin{eqnarray}
	\left(\Gamma_{AA}^{\Omega}\right)_{\alpha\beta}^{mn}\left(p,q\right) & = & \tilde{\Omega}_{\mu}\left(p+q\right)igv\varepsilon^{3nm}\left[\left(p_{\alpha}+2q_{\alpha}\right)\delta_{\beta\mu}-\left(2p_{\beta}+q_{\beta}\right)\delta_{\alpha\mu}\right.\nonumber \\
	&  & \left.-\delta_{\alpha\beta}\left(q_{\mu}-p_{\mu}\right)\right]
\end{eqnarray}

\begin{itemize}
	\item Three vectors vertex
\end{itemize}
\begin{eqnarray}
	\left(\Gamma_{AAA}^{\Omega}\right)_{\alpha\beta\gamma}^{mnp}\left(p,q,r\right) & = & g^{2}v\tilde{\Omega}_{\mu}\left(p+q+r\right)\left[\left(\varepsilon^{3bm}\varepsilon^{bnp}+\varepsilon^{3bn}\varepsilon^{bmp}\right)\delta_{\alpha\beta}\delta_{\gamma\mu}\right.\nonumber \\
	&  & +\left(\varepsilon^{3bm}\varepsilon^{bpn}+\varepsilon^{3bp}\varepsilon^{bmn}\right)\delta_{\alpha\gamma}\delta_{\beta\mu}\nonumber \\
	&  & \left.+\left(\varepsilon^{3bn}\varepsilon^{bpm}+\varepsilon^{3bp}\varepsilon^{bnm}\right)\delta_{\beta\gamma}\delta_{\alpha\mu}\right]
\end{eqnarray}

\begin{itemize}
	\item Vector-scalar vertex
\end{itemize}
\begin{eqnarray}
	\left(\Gamma_{AH}^{\Omega}\right)_{\alpha}^{mn}\left(p,q\right) & = & \delta^{mn}\left(p^{2}\delta_{\alpha\mu}-p_{\alpha}p_{\mu}\right)\tilde{\Omega}_{\mu}\left(p+q\right)
\end{eqnarray}

\begin{itemize}
	\item Vector-vector-scalar vertex
\end{itemize}
\begin{eqnarray}
	\left(\Gamma_{AAH}^{\Omega}\right)_{\alpha\beta}^{mnp}\left(p,q,r\right) & = & \tilde{\Omega}_{\mu}\left(p+q+r\right)ig\varepsilon^{pnm}\left[\left(p_{\alpha}+2q_{\alpha}\right)\delta_{\beta\mu}-\left(2p_{\beta}+q_{\beta}\right)\delta_{\alpha\mu}\right.\nonumber \\
	&  & \left.-\delta_{\alpha\beta}\left(q_{\mu}-p_{\mu}\right)\right]
\end{eqnarray}

\begin{itemize}
	\item Vector-vector-vector-scalar vertex
\end{itemize}
\begin{eqnarray}
	\left(\Gamma_{AAAH}^{\Omega}\right)_{\alpha\beta\gamma}^{mnpq}\left(p,q,r,s\right) & = & g^{2}\tilde{\Omega}_{\mu}\left(p+q+r+s\right)\left[\left(\varepsilon^{qbm}\varepsilon^{bnp}+\varepsilon^{qbn}\varepsilon^{bmp}\right)\delta_{\alpha\beta}\delta_{\mu\gamma}\right.\nonumber \\
	&  & +\left(\varepsilon^{qbm}\varepsilon^{bpn}+\varepsilon^{qbp}\varepsilon^{bmn}\right)\delta_{\alpha\gamma}\delta_{\mu\beta}\nonumber \\
	&  & \left.+\left(\varepsilon^{qbn}\varepsilon^{bpm}+\varepsilon^{qbp}\varepsilon^{bnm}\right)\delta_{\beta\gamma}\delta_{\mu\beta}\right]
\end{eqnarray}

\section{One-loop corrections for $\left\langle O_{\mu}\left(x\right)O_{\nu}\left(y\right)\right\rangle $}\label{One-loop corrections for}

The connected two-point Green's function of $O_{\mu}$ is given by
\begin{equation}
	\langle O_\mu(x)O_\nu(y)\rangle = - \left.\frac{\delta^2 W[J_{\varphi},\Omega]}{\delta \Omega_\mu(x) \Omega_\nu(y)}\right|_{J_{\varphi}=\Omega=0},
\end{equation}
where generating functional of connected Green's functions is defined as
\begin{equation}
	W[J_{\varphi},\Omega]=-\ln N \int D\varphi e^{-(S+\int d^4x J_{\varphi}\cdot \varphi+\int d^4x \Omega_\mu O_{\mu})}.
\end{equation}
In this work, we adopted the standard approach for computing Green’s functions of composite operators, in which the operator is coupled to an external source. More recently, we presented an alternative but equivalent method based on the introduction of an auxiliary field, see Ref.~\cite{Peruzzo:2025awi}.. \par

The one-loop Feynman diagram contributing to $\left\langle O_{\mu}\left(x\right)O_{\nu}\left(y\right)\right\rangle $
are shown in Figure 2. Below, we present each one of the corrections
in the order they appear in this figure. All the results will be expressed
as a linear combination of the master integrals (\ref{eq:xi_master})
and
\begin{eqnarray}
	\eta\left(p^{2},m_{1}^{2},m_{2}^{2}\right) & = & \int\frac{d^{d}k}{\left(2\pi\right)^{d}}\frac{1}{k^{2}+m_{1}^{2}}\frac{1}{\left(p-k\right)^{2}+m_{2}^{2}}\nonumber \\
	& = & \frac{1}{\left(4\pi\right)^{\frac{d}{2}}}\Gamma\left(2-\frac{d}{2}\right)\int_{0}^{1}dx\left[p^{2}x\left(1-x\right)+m_{1}^{2}x+m_{2}^{2}\left(1-x\right)\right]^{\frac{d}{2}-2}.
\end{eqnarray}

\begin{figure}
	\begin{centering}
		\includegraphics[scale=0.35]{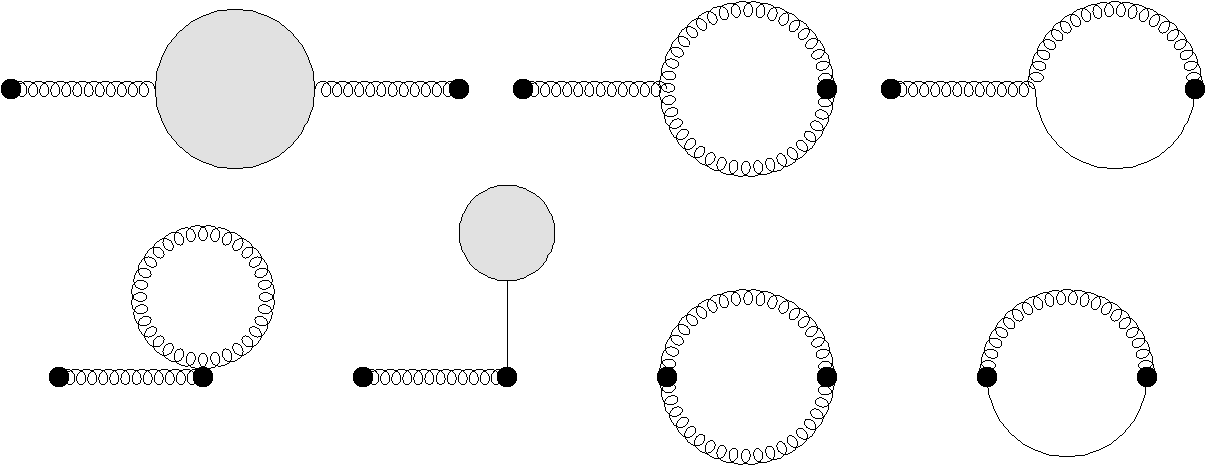}
		\par\end{centering}
	\caption{One-loop Feynman diagrams contributing to $\left\langle O_{\mu}\left(x\right)O_{\nu}\left(y\right)\right\rangle $.
		The full two-point function $\left\langle A_{\mu}^{a}\left(x\right)A_{\nu}^{b}\left(y\right)\right\rangle $
		and the full one-point function $\left\langle h\left(x\right)\right\rangle $
		are represented by shaded circles. }\label{fig:One-loop-Feynman-diagrams}
\end{figure}

\begin{itemize}
	\item First diagram
\end{itemize}
The first diagram contains all one-loop corrections to $\left\langle A_{\mu}^{a}\left(x\right)A_{\nu}^{b}\left(y\right)\right\rangle $,
which are shown in Figure 3. We indicate each diagram with a Latin
letter as follows:

a)
\begin{eqnarray}
	\left(I^{a}\right)_{\mu\nu} & = & \frac{1}{2}v\delta^{a3}\left(\delta_{\alpha\mu}p^{2}-p_{\alpha}p_{\mu}\right)\left(\Delta_{AA}\right)_{\alpha\beta}^{am}\left(p\right)\int\frac{d^{d}k}{\left(2\pi\right)}\left(\Gamma_{AAA}\right)_{\beta\gamma\delta}^{mnp}\left(p,-k,-p+k\right)\left(\Delta_{AA}\right)_{\gamma\rho}^{nr}\left(k\right)\times\nonumber \\
	&  & \times\left(\Gamma_{AAA}\right)_{\rho\sigma\tau}^{rst}\left(k,p-k,-p\right)\left(\Delta_{AA}\right)_{\delta\sigma}^{ps}\left(p-k\right)\left(\Delta_{AA}\right)_{\tau\lambda}^{tu}\left(p\right)v\delta^{u3}\left(\delta_{\lambda\nu}p^{2}-p_{\lambda}p_{\nu}\right)
\end{eqnarray}
which corresponds to
\begin{eqnarray*}
	\left(I^{a}\right)_{\mu\nu} & = & i^{2}v^{2}g^{2}T_{\mu\beta}\left(p\right)\Pi_{\beta\tau}\left(p\right)T_{\tau\nu}\left(p\right)
\end{eqnarray*}
where
\begin{eqnarray}
	\Pi_{\beta\tau}\left(p\right) & = & \int\frac{d^{d}k}{\left(2\pi\right)}\left[\left(-2k_{\beta}+p_{\beta}\right)\delta_{\gamma\delta}+\left(-2p_{\gamma}+k_{\gamma}\right)\delta_{\delta\beta}+\left(p_{\delta}+k_{\delta}\right)\delta_{\beta\gamma}\right]\frac{T_{\gamma\rho}\left(k\right)}{k^{2}+m^{2}}\times\nonumber \\
	&  & \times\left[\left(2p_{\rho}-k_{\rho}\right)\delta_{\sigma\tau}+\left(-p_{\sigma}-k_{\sigma}\right)\delta_{\tau\rho}+\left(2k_{\tau}-p_{\tau}\right)\delta_{\rho\sigma}\right]\frac{T_{\delta\sigma}\left(p-k\right)}{\left(p-k\right)^{2}+m^{2}}.\label{eq:PI_def}
\end{eqnarray}

b) 

\begin{eqnarray}
	\left(I^{b}\right)_{\mu\nu} & = & \left(-\right)v\delta^{a3}\left(\delta_{\alpha\mu}p^{2}-p_{\alpha}p_{\mu}\right)\left(\Delta_{AA}\right)_{\alpha\beta}^{am}\left(p\right)\int\frac{d^{d}k}{\left(2\pi\right)^{d}}\left(\Gamma_{A\overline{c}c}\right)_{\beta}^{mpn}\left(p,-p+k,-k\right)\left(\Delta_{c\overline{c}}\right)^{nr}\left(k\right)\nonumber \\
	&  & \left(\Gamma_{A\overline{c}c}\right)_{\tau}^{trs}\left(-p,k,p-k\right)\left(\Delta_{c\overline{c}}\right)^{sp}\left(p-k\right)\left(\Delta_{AA}\right)_{\tau\lambda}^{tu}\left(p\right)v\delta^{u3}\left(p^{2}\delta_{\lambda\nu}-p_{\lambda}p_{\nu}\right)
\end{eqnarray}
which corresponds to

\begin{eqnarray}
	\left(I^{b}\right)_{\mu\nu} & = & T_{\mu\nu}\left(p\right)\frac{p^{2}m^{2}}{2\left(d-1\right)}\eta\left(p^{2},0,0\right).
\end{eqnarray}

c) 

\begin{eqnarray}
	\left(I^{c}\right)_{\mu\nu} & = & \frac{1}{2}v\delta^{a3}\left(\delta_{\alpha\mu}p^{2}-p_{\alpha}p_{\mu}\right)\left(\Delta_{AA}\right)_{\alpha\beta}^{am}\left(p\right)\int\frac{d^{d}k}{\left(2\pi\right)^{d}}\left(\Gamma_{AHH}\right)_{\beta}^{mpn}\left(p,-p+k,-k\right)\left(\Delta_{HH}\right)^{nr}\left(k\right)\nonumber \\
	&  & \left(\Gamma_{AHH}\right)_{\tau}^{trs}\left(-p,k,p-k\right)\left(\Delta_{HH}\right)^{sp}\left(p-k\right)\left(\Delta_{AA}\right)_{\tau\lambda}^{tb}\left(p\right)v\delta^{b3}\left(p^{2}\delta_{\lambda\nu}-p_{\lambda}p_{\nu}\right)
\end{eqnarray}
which corresponds to
\begin{eqnarray}
	\left(I^{c}\right)_{\mu\nu} & = & -T_{\mu\nu}\left(p\right)\frac{p^{2}m^{2}}{\left(d-1\right)}\eta\left(p^{2},0,0\right).
\end{eqnarray}

d)

\begin{eqnarray}
	\left(I^{d}\right)_{\mu\nu} & = & v\delta^{3a}\left(\delta_{\mu\alpha}p^{2}-p_{\mu}p_{\alpha}\right)\left(\Delta_{AA}\right)_{\alpha\beta}^{am}\left(p\right)\int\frac{d^{d}k}{\left(2\pi\right)^{d}}\left(\Gamma_{AAH}\right)_{\beta\gamma}^{mnp}\left(p,-k,-p+k\right)\left(\Delta_{AA}\right)_{\gamma\delta}^{nr}\left(k\right)\nonumber \\
	&  & \left(\Gamma_{AAH}\right)_{\delta\tau}^{rts}\left(k,-p,p-k\right)\left(\Delta_{HH}\right)^{ps}\left(p-k\right)\left(\Delta_{AA}\right)_{\tau\lambda}^{tb}\left(p\right)v\delta^{3b}\left(p^{2}\delta_{\lambda\nu}-p_{\lambda}p_{\nu}\right)
\end{eqnarray}
which corresponds to
\begin{eqnarray}
	\left(I^{d}\right)_{\mu\nu} & = & 2v^{4}g^{4}T_{\mu\beta}\left(p\right)T_{\tau\nu}\left(p\right)\int\frac{d^{d}k}{\left(2\pi\right)^{d}}\frac{T_{\beta\tau}\left(k\right)}{k^{2}+m^{2}}\frac{1}{\left(p-k\right)^{2}}.
\end{eqnarray}

e)

\begin{eqnarray}
	\left(I^{e}\right)_{\mu\nu} & = & \frac{1}{2}v\delta^{3a}\left(p^{2}\delta_{\mu\alpha}-p_{\mu}p_{\alpha}\right)\left(\Delta_{AA}\right)_{\alpha\beta}^{am}\left(p\right)\int\frac{d^{d}k}{\left(2\pi\right)^{d}}\left(\Gamma_{AAAA}\right)_{\alpha\beta\gamma\delta}^{mnpq}\left(p,k,-k,-p\right)\nonumber \\
	&  & \left(\Delta_{AA}\right)_{\gamma\delta}^{np}\left(k\right)\left(\Delta_{AA}\right)_{\epsilon\lambda}^{qb}\left(p\right)v\delta^{3b}\left(p^{2}\delta_{\lambda\nu}-p_{\lambda}p_{\nu}\right)
\end{eqnarray}
which corresponds to
\begin{eqnarray}
	\left(I^{e}\right)_{\mu\nu} & = & -2g^{2}v^{2}T_{\mu\nu}\left(p\right)\frac{\left(d-1\right)^{2}}{d}\chi\left(m^{2}\right).
\end{eqnarray}

f)
\begin{eqnarray}
	\left(I^{f}\right)_{\mu\nu} & = & \frac{1}{2}v\delta^{3a}\left(p^{2}\delta_{\mu\tau}-p_{\mu}p_{\tau}\right)\left(\Delta_{AA}\right)_{\tau\alpha}^{am}\left(p\right)\int\frac{d^{d}k}{\left(2\pi\right)^{d}}\left(\Gamma_{AAHH}\right)_{\alpha\beta}^{mnpq}\left(p,-k,k,-p\right)\nonumber \\
	&  & \left(\Delta_{HH}\right)^{pq}\left(k\right)\left(\Delta_{AA}\right)_{\beta\lambda}^{nb}\left(p\right)v\delta^{b3}\left(p^{2}\delta_{\lambda\nu}-p_{\lambda}p_{\nu}\right)
\end{eqnarray}
which vanishes due to the massless nature of the would-be Goldstone
fields, \emph{i.e.},
\begin{eqnarray}
	\left(I^{f}\right)_{\mu\nu} & = & 0.
\end{eqnarray}

g) This diagram contains the full one-point function $\left\langle H^{a}\left(0\right)\right\rangle $.
Nevertheless, due to the particular group-tensor structure of $\left(\Gamma_{AAH}\right)_{\mu\nu}^{abc}\left(p,q,r\right)$,
it vanishes identically, \emph{i.e.},
\begin{eqnarray}
	\left(I^{g}\right)_{\mu\nu} & = & 0.
\end{eqnarray}

\begin{figure}
	\begin{centering}
		\includegraphics[scale=0.35]{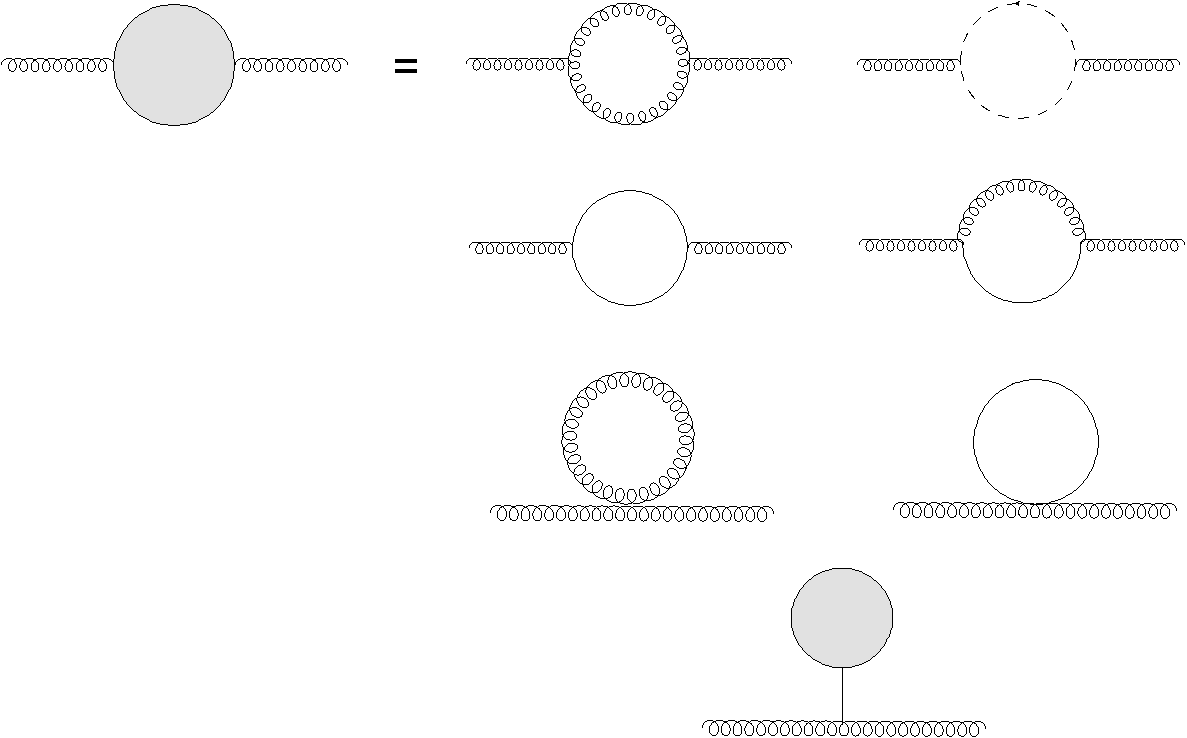}
		\par\end{centering}
	\caption{One-loop Feynman diagrams contributing to $\left\langle A_{\mu}^{a}\left(x\right)A_{\nu}^{b}\left(y\right)\right\rangle $.
		The full one-point function $\left\langle h\left(x\right)\right\rangle $
		is represented by a shaded circle.}
	
\end{figure}

\begin{itemize}
	\item Second diagram
\end{itemize}
\begin{eqnarray}
	\left(II\right)_{\mu\nu} & = & \frac{1}{2}v\delta^{a3}\left(\delta_{\alpha\mu}p^{2}-p_{\alpha}p_{\mu}\right)\left(\Delta_{AA}\right)_{\alpha\beta}^{am}\left(p\right)\nonumber \\
	&  & \int\frac{d^{d}k}{\left(2\pi\right)}\left(\Gamma_{AAA}\right)_{\beta\gamma\delta}^{mnp}\left(p,-k,-p+k\right)\left(\Delta_{AA}\right)_{\gamma\rho}^{nr}\left(k\right)\left(\Gamma_{AA}^{\Omega}\right)_{\rho\sigma\nu}^{rs3}\left(k,p-k\right)\left(\Delta_{AA}\right)_{\delta\sigma}^{ps}\left(p-k\right)
\end{eqnarray}
which corresponds to
\begin{eqnarray*}
	\left(II\right)_{\mu\nu} & = & -g^{2}v^{2}i^{2}T_{\mu\beta}\left(p\right)\Pi_{\beta\nu}\left(p\right),
\end{eqnarray*}
where the tensor $\Pi_{\beta\nu}\left(p\right)$ is defined in Eq.
(\ref{eq:PI_def}). There is also a mirror diagram whose result is
\begin{eqnarray*}
	\left(II^{mirror}\right)_{\mu\nu} & = & -g^{2}v^{2}i^{2}\Pi_{\mu\beta}\left(p\right)T_{\beta\nu}\left(p\right),
\end{eqnarray*}

\begin{itemize}
	\item Third diagram
\end{itemize}
\begin{eqnarray}
	\left(III\right)_{\mu\nu} & = & v\delta^{3a}\left(p^{2}\delta_{\mu\alpha}-p_{\mu}p_{\alpha}\right)\left(\Delta_{AA}\right)_{\alpha\beta}^{am}\left(p\right)\int\frac{d^{d}k}{\left(2\pi\right)^{d}}\left(\Gamma_{AAH}\right)_{\beta\gamma}^{mnp}\left(p,-k,-p+k\right)\left(\Delta_{AA}\right)_{\gamma\delta}^{nr}\left(k\right)\times\nonumber \\
	&  & \times\delta^{rs}\left(k^{2}\delta_{\delta\nu}-k_{\delta}k_{\nu}\right)\left(\Delta_{HH}\right)^{ps}\left(p-k\right)
\end{eqnarray}
which results in
\begin{eqnarray}
	\left(III\right)_{\mu\nu} & = & 2v^{2}g^{2}T_{\mu\beta}\left(p\right)\int\frac{d^{d}k}{\left(2\pi\right)^{d}}\frac{T_{\beta\nu}\left(k\right)}{k^{2}+m^{2}}\frac{1}{\left(p-k\right)^{2}}k^{2}.
\end{eqnarray}
There is also a mirror diagram that we have to take into account whose
results is
\begin{eqnarray}
	\left(III^{mirror}\right)_{\mu\nu} & = & 2v^{2}g^{2}T_{\nu\beta}\left(p\right)\int\frac{d^{d}k}{\left(2\pi\right)^{d}}\frac{T_{\beta\mu}\left(k\right)}{k^{2}+m^{2}}\frac{1}{\left(p-k\right)^{2}}k^{2}.
\end{eqnarray}

\begin{itemize}
	\item Fourth diagram
\end{itemize}
\begin{eqnarray}
	\left(IV\right)_{\mu\nu} & = & \frac{1}{2}v\delta^{3a}\left(p^{2}\delta_{\mu\tau}-p_{\mu}p_{\tau}\right)\left(\Delta_{AA}\right)_{\tau\alpha}^{am}\left(p\right)\nonumber \\
	&  & \int\frac{d^{d}k}{\left(2\pi\right)^{d}}g^{2}v\left[\left(\varepsilon^{3bm}\varepsilon^{bnp}+\varepsilon^{3bn}\varepsilon^{bmp}\right)\delta_{\alpha\beta}\delta_{\gamma\nu}\right.\nonumber \\
	&  & +\left(\varepsilon^{3bm}\varepsilon^{bpn}+\varepsilon^{3bp}\varepsilon^{bmn}\right)\delta_{\alpha\gamma}\delta_{\beta\nu}\nonumber \\
	&  & \left.+\left(\varepsilon^{3bn}\varepsilon^{bpm}+\varepsilon^{3bp}\varepsilon^{bnm}\right)\delta_{\beta\gamma}\delta_{\alpha\nu}\right]\left(\Delta_{AA}\right)_{\beta\gamma}^{np}\left(k\right)
\end{eqnarray}
which results in
\begin{eqnarray}
	\left(IV\right)_{\mu\nu} & = & 2g^{2}v^{2}T_{\mu\nu}\left(p\right)\frac{\left(d-1\right)^{2}}{d}\chi\left(m^{2}\right).
\end{eqnarray}
There is also a mirror diagram whose result is
\begin{eqnarray}
	\left(IV^{mirror}\right)_{\mu\nu} & = & 2g^{2}v^{2}T_{\mu\nu}\left(p\right)\frac{\left(d-1\right)^{2}}{d}\chi\left(m^{2}\right).
\end{eqnarray}

\begin{itemize}
	\item Fifth diagram
\end{itemize}
\begin{eqnarray*}
	\left(V\right)_{\mu\nu} & = & v\delta^{a3}\left(p^{2}\delta_{\mu\tau}-p_{\mu}p_{\tau}\right)\left(\Delta_{AA}\right)_{\tau\alpha}^{am}\left(p\right)\delta^{mn}\left(p^{2}\delta_{\alpha\nu}-p_{\alpha}p_{\nu}\right)\Delta^{nq}\left(0\right)\left\langle H^{q}\left(0\right)\right\rangle _{1PI}
\end{eqnarray*}
where $\left\langle H^{q}\left(0\right)\right\rangle _{1PI}$ is the
one-point 1PI function of $H^{q}\left(x\right)$. After we contract
all the tensor-group indices, it follows that
\begin{eqnarray}
	\left(V\right)_{\mu\nu} & = & T_{\mu\nu}\left(p\right)vp^{2}\frac{1}{m_{h}^{2}}\left\langle H^{3}\left(0\right)\right\rangle _{1PI}.
\end{eqnarray}
All the diagrams that contribute to $\left\langle H^{3}\left(0\right)\right\rangle _{1PI}$
are shown in Figure 3. The first diagram of Figure 3 is
\begin{eqnarray}
	\left\langle H^{3}\left(0\right)\right\rangle _{1PI}^{\left(a\right)} & = & -\frac{1}{2}\lambda v\left(\delta^{33}\delta^{np}+\delta^{3p}\delta^{n3}+\delta^{3n}\delta^{p3}\right)\int\frac{d^{d}k}{\left(2\pi\right)^{d}}\left(\Delta_{HH}\right)^{np}\left(k\right),
\end{eqnarray}
which results in
\begin{eqnarray}
	\left\langle H^{3}\left(0\right)\right\rangle _{1PI}^{\left(a\right)} & = & -\frac{3}{2}\lambda v\chi\left(m_{h}^{2}\right),
\end{eqnarray}
whereas the second one is
\begin{eqnarray}
	\left\langle H^{3}\left(0\right)\right\rangle _{1PI}^{\left(b\right)} & = & -\frac{1}{2}vg^{2}\left(\varepsilon^{a33}\varepsilon^{apn}+\varepsilon^{a3n}\varepsilon^{ap3}\right)\delta_{\alpha\beta}\int\frac{d^{d}k}{\left(2\pi\right)^{d}}\left(\Delta_{AA}\right)_{\alpha\beta}^{np}\left(k\right)
\end{eqnarray}
whose result after we perform the integration is
\begin{eqnarray}
	\left\langle H^{3}\left(0\right)\right\rangle _{1PI}^{\left(b\right)} & = & vg^{2}\left(d-1\right)\chi\left(m^{2}\right).
\end{eqnarray}
Therefore, we obtain the following result for the one-point function
\begin{eqnarray*}
	\left\langle H^{3}\left(0\right)\right\rangle _{1PI} & = & -\frac{3}{2}\lambda v\chi\left(m_{h}^{2}\right)+vg^{2}\left(d-1\right)\chi\left(m^{2}\right).
\end{eqnarray*}
Consequently, it follows that
\begin{eqnarray}
	\left(V\right)_{\mu\nu} & = & T_{\mu\nu}\left(p\right)p^{2}\left[-\frac{3}{2}\chi\left(m_{h}^{2}\right)+\frac{m^{2}}{m_{h}^{2}}\left(d-1\right)\chi\left(m^{2}\right)\right].
\end{eqnarray}
There is also a mirror diagram whose result is the same
\begin{eqnarray}
	\left(V^{mirror}\right)_{\mu\nu} & = & T_{\mu\nu}\left(p\right)p^{2}\left[-\frac{3}{2}\chi\left(m_{h}^{2}\right)+\frac{m^{2}}{m_{h}^{2}}\left(d-1\right)\chi\left(m^{2}\right)\right].
\end{eqnarray}

\begin{figure}
	\begin{centering}
		\includegraphics[scale=0.3]{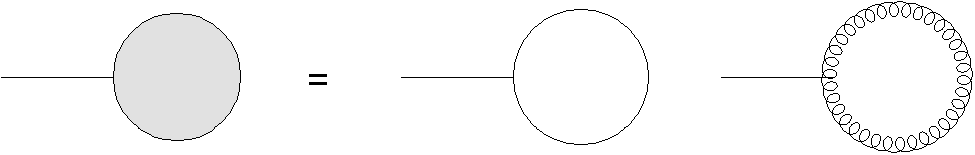}
		\par\end{centering}
	\caption{Tadpole diagrams that contribute to the one-point 1PI function $\left\langle H^{3}\left(0\right)\right\rangle _{1PI}$.}
\end{figure}

\begin{itemize}
	\item Sixth diagram
\end{itemize}
\begin{eqnarray}
	\left(VI\right)_{\mu\nu} & = & \frac{1}{2}igv\varepsilon^{3pn}\int\frac{d^{d}k}{\left(2\pi\right)^{d}}\left[\left(-k_{\gamma}-2\left(p-k\right)_{\gamma}\right)\delta_{\delta\mu}-\left(-2k_{\delta}-\left(p-k\right)_{\delta}\right)\delta_{\gamma\mu}-\delta_{\gamma\delta}\left(-\left(p-k\right)_{\mu}+k_{\mu}\right)\right]\left(\Delta_{AA}\right)_{\gamma\rho}^{nr}\left(k\right)\nonumber \\
	&  & igv\varepsilon^{3sr}\left[\left(k_{\rho}+2\left(p-k\right)_{\rho}\right)\delta_{\sigma\nu}-\left(2k_{\sigma}+\left(p-k\right)_{\sigma}\right)\delta_{\rho\nu}-\delta_{\rho\sigma}\left(\left(p-k\right)_{\nu}-k_{\nu}\right)\right]\left(\Delta_{AA}\right)_{\delta\sigma}^{ps}\left(p-k\right)
\end{eqnarray}
which corresponds to
\begin{eqnarray}
	\left(VI\right)_{\mu\nu} & = & i^{2}g^{2}v^{2}\Pi_{\mu\nu}\left(p\right),
\end{eqnarray}
where the tensor $\Pi_{\mu\nu}\left(p\right)$ is defined in Eq. (\ref{eq:PI_def}).
\begin{itemize}
	\item Seventh diagram
\end{itemize}
\begin{eqnarray}
	\left(VII\right)_{\mu\nu} & = & \int\frac{d^{d}k}{\left(2\pi\right)^{d}}\delta^{np}\left(k^{2}\delta_{\mu\gamma}-k_{\mu}k_{\gamma}\right)\left(\Delta_{AA}\right)_{\gamma\delta}^{nr}\left(k\right)\delta^{rs}\left(k^{2}\delta_{\delta\nu}-k_{\delta}k_{\nu}\right)\left(\Delta_{HH}\right)^{ps}\left(p-k\right)
\end{eqnarray}
which corresponds to
\begin{eqnarray}
	\left(VII\right)_{\mu\nu} & = & \int\frac{d^{d}k}{\left(2\pi\right)^{d}}\frac{T_{\mu\nu}\left(k\right)}{k^{2}}\frac{1}{\left(p-k\right)^{2}+m_{h}^{2}}k^{4}+2\int\frac{d^{d}k}{\left(2\pi\right)^{d}}\frac{T_{\mu\nu}\left(k\right)}{k^{2}+m^{2}}\frac{1}{\left(p-k\right)^{2}}k^{4}.
\end{eqnarray}

For some diagrams, we did not present the expression after performing
the loop integration because it is more convenient to
first add up all diagrams with the same loop lines. Adding all diagrams
with one vector and one scalar loop lines, we obtain

\begin{gather}
	\left(I^{d}\right)_{\mu\nu}+\left(III\right)_{\mu\nu}+\left(III^{mirror}\right)_{\mu\nu}+\left(VII\right)_{\mu\nu}\nonumber \\
	=\frac{p^{2}m^{2}}{2\left(d-1\right)}T_{\mu\nu}\left(p\right)\eta\left(p^{2},0,0\right)+2m^{4}L_{\mu\nu}\left(p\right)\left[\left(1+\frac{\left(p^{2}-m^{2}\right)^{2}}{4p^{2}m^{2}}\right)\eta\left(p^{2},m^{2},0\right)-\frac{p^{2}-m^{2}}{4p^{2}m^{2}}\chi\left(m^{2}\right)\right]\nonumber \\
	+\delta_{\mu\nu}\left(-m_{h}^{2}+p^{2}\right)\chi\left(m_{h}^{2}\right)-\left(-\frac{m_{h}^{2}}{d}\delta_{\mu\nu}+p_{\mu}p_{\nu}\right)\chi\left(m_{h}^{2}\right).
\end{gather}

For diagrams with two vector loop lines, it follows that
\begin{gather}
	\left(I^{a}\right)_{\mu\nu}+\left(II\right)_{\mu\nu}+\left(II^{mirror}\right)_{\mu\nu}+\left(VI\right)_{\mu\nu}\nonumber \\
	=i^{2}v^{2}g^{2}\left(T_{\mu\beta}\left(p\right)\Pi_{\beta\tau}\left(p\right)T_{\tau\nu}\left(p\right)-T_{\mu\beta}\left(p\right)\Pi_{\beta\nu}\left(p\right)-\Pi_{\mu\beta}\left(p\right)T_{\beta\nu}\left(p\right)+\Pi_{\mu\nu}\left(p\right)\right).\label{eq:sum_vector_loop}
\end{gather}
From the Lorentz invariance, we know that
\begin{eqnarray}
	\Pi_{\mu\nu}\left(p\right) & = & \frac{T_{\alpha\beta}\Pi_{\alpha\beta}}{\left(d-1\right)}T_{\mu\nu}\left(p\right)+L_{\alpha\beta}\Pi_{\alpha\beta}L_{\mu\nu}\left(p\right).
\end{eqnarray}
Therefore, we obtain that the sum (\ref{eq:sum_vector_loop}) is longitudinal,
\emph{i.e.},
\begin{gather}
	\left(I^{a}\right)_{\mu\nu}+\left(II\right)_{\mu\nu}+\left(II^{mirror}\right)_{\mu\nu}+\left(VI\right)_{\mu\nu}=i^{2}v^{2}g^{2}L_{\alpha\beta}\Pi_{\alpha\beta}L_{\mu\nu}\left(p\right).
\end{gather}
Explicitly computing the longitudinal part of $\Pi_{\mu\nu}\left(p\right)$,
we obtain
\begin{eqnarray*}
	L_{\alpha\beta}\Pi_{\alpha\beta}\left(p\right) & = & 2m^{2}\left[1+\frac{\left(p^{2}-m^{2}\right)^{2}}{4p^{2}m^{2}}\right]\eta\left(p^{2},m^{2},0\right)-\left[2\left(d-2\right)+\frac{2}{d}+\frac{\left(p^{2}-m^{2}\right)}{2p^{2}}\right]\chi\left(m^{2}\right).
\end{eqnarray*}
Thus, it yields
\begin{gather*}
	\left(I^{a}\right)_{\mu\nu}+\left(II\right)_{\mu\nu}+\left(II^{mirror}\right)_{\mu\nu}+\left(VI\right)_{\mu\nu}\\
	=L_{\mu\nu}\left(p\right)\left\{ -2m^{4}\left[1+\frac{\left(p^{2}-m^{2}\right)^{2}}{4p^{2}m^{2}}\right]\eta\left(p^{2},m^{2},0\right)+m^{2}\left[2\left(d-2\right)+\frac{2}{d}+\frac{\left(p^{2}-m^{2}\right)}{2p^{2}}\right]\chi\left(m^{2}\right)\right\} .
\end{gather*}

Finally, summing all contributions, we obtain
\begin{eqnarray}
	\left\langle O_{\mu}\left(p\right)O_{\nu}\left(-p\right)\right\rangle _{\textrm{one-loop}} & = & \delta_{\mu\nu}\left(-m_{h}^{2}+p^{2}\right)\chi\left(m_{h}^{2}\right)-\left(-\frac{m_{h}^{2}}{d}\delta_{\mu\nu}+p_{\mu}p_{\nu}\right)\chi\left(m_{h}^{2}\right)+2m^{2}\delta_{\mu\nu}\frac{\left(d-1\right)^{2}}{d}\chi\left(m^{2}\right)\nonumber \\
	&  & +2\left(p^{2}\delta_{\mu\nu}-p_{\mu}p_{\nu}\right)\left[-\frac{3}{2}\chi\left(m_{h}^{2}\right)+\frac{m^{2}}{m_{h}^{2}}\left(d-1\right)\chi\left(m^{2}\right)\right],
\end{eqnarray}
which is a local function of the external momentum $p$, as claimed.


\end{document}